\documentclass[fleqn,twoside]{article}
\usepackage{espcrc2}
\usepackage{psfig}
\newcommand{\beeq}{\begin{equation}}
\newcommand{\eneq}{\end{equation}}
\newcommand{\beeqa}{\begin{eqnarray*}}
\newcommand{\eneqa}{\end{eqnarray*}}

\newcommand{\bma}{\begin{displaymath}}
\newcommand{\ema}{\end{displaymath}}

\title{SUSY Ward identities in 1-loop perturbation theory}

\author{Federico Farchioni $\rm^\dagger$ \address{Deutsches Elektronen-Synchrotron, DESY,
        Notkestr.~85, D-22603 Hamburg, Germany},
        Alessandra Feo\address{Institut f\"ur Theoretische Physik,
                     Universit\"at M\"unster, \\
                     Wilhelm-Klemm-Str.~9, D-48149 M\"unster, Germany}
        \thanks{\protect{Talk given by Alessandra Feo.\hfill\hfill\break
         $\rm^\dagger$ Present Address: 
           Institut f\"ur Theoretische Physik,
                     Universit\"at M\"unster,
                     Wilhelm-Klemm-Str.~9, D-48149 M\"unster, Germany }},
        Tobias Galla\address{Department of Physics, University of Oxford,
                     1 Keble Road, Oxford OX1 3NP, UK},
        Claus Gebert$\rm^a$,
        Robert Kirchner$\rm^a$, \\
        Istv\'an Montvay$\rm^a$,
        Gernot M\"unster$\rm^b$,
        Roland Peetz $\rm^b$,
        Anastassios Vladikas\address{INFN, Sezione di Roma 2,
                  Universit\'a di Roma ``Tor Vergata'', I-00133 Rome, Italy}.
         \\[0.5em]
        DESY-M\"unster-Roma Collaboration \\ }

\begin{document}


\begin{abstract}

We present preliminary results of a study of the 
supersymmetric (SUSY) Ward identities (WIs) for the $N=1$ SU$(2)$ 
SUSY Yang-Mills theory in the context of one-loop lattice perturbation theory.
The supersymmetry on the lattice is explicitly broken by  
the gluino mass and the lattice artifacts.
However, the renormalization of the supercurrent can be 
carried out in a scheme that restores the nominal continuum WIs. 
The perturbative calculation of the renormalization constants and
mixing coefficients for the local supercurrent is presented.

\end{abstract}

\maketitle

\section{INTRODUCTION}
\vspace{-0.1 cm}

SUSY gauge theories present different non-perturbative aspects
which are the object of current research,
for example the possible mechanisms for dynamical supersymmetry breaking.
The simplest SUSY gauge theory is the $N=1$ SUSY Yang-Mills theory. 
For SU$(N_c)$ it has $(N_c^2 - 1)$ gluons 
and the same number of massless Majorana fermions (gluinos)
in the adjoint representation of the color group.

To formulate supersymmetry on the lattice we follow the ideas of 
Curci and Veneziano \cite{curci}. They adopt the Wilson
formulation for the $N=1$ SUSY Yang-Mills theory. 
Supersymmetry is broken explicitly by the Wilson
term and the finite lattice spacing. In addition, a soft breaking 
due to the introduction of the gluino mass is present. 
It is proposed that
supersymmetry can be recovered in the continuum limit by tuning the bare 
gauge coupling $g$ and the gluino mass $m_{\tilde{g}}$ to the SUSY point,
at $m_{\tilde{g}} =0$, which also coincides with the chiral point.
In previous publications \cite{campos}, and references therein, we have 
investigated these issues.

Another independent way to study the SUSY (chiral) limit is by means of
the SUSY WIs. On the lattice they contain explicit SUSY
breaking terms.
In this framework, the SUSY limit is defined to be the point in parameter
space where these breaking terms vanish and the SUSY WIs take their
continuum form.

Our collaboration is currently focussing on the study 
of the SUSY WIs on the lattice, either with Monte Carlo methods
\cite{federico} or by a perturbative calculation
of the renormalization constants and mixing coefficients 
of the lattice supercurrent. 
For a different perturbative approach to SUSY WIs see \cite{taniguchi}.

\vspace{-0.2 cm}
\section{SUSY WARD IDENTITIES ON THE LATTICE}
\vspace{-0.1 cm}

In the Wilson formulation of the $N=1$ SUSY Yang-Mills theory 
\cite{curci} the gluonic part of the action is the standard plaquette action
while the fermionic part reads 
\vspace{-0.1 cm}
\beeqa
S_f &= & \mbox{Tr} \left\{ 
\frac{1}{2a} \Bigg( \bar\lambda(x)( \gamma_\mu - r) U_\mu^\dagger(x)
\lambda(x + a \hat\mu) U_\mu(x) \right. \\[-1mm]
&&  \left. - \bar\lambda(x + a \hat\mu) (\gamma_\mu + r) U_\mu(x)
\lambda(x) U_\mu^\dagger(x) \Bigg) \right.  \\[-1mm]
&& \left. + \left(m_0 + \frac{4r}{a} \right) \bar\lambda(x)
\lambda(x) \right\} \, .
\eneqa
Supersymmetry is not realized on the lattice. One can still
define transformations that reduce to the continuum SUSY ones
in the limit $a \to 0 $. A possible choice is
\beeqa
\delta U_\mu(x) & = & - a 
g U_\mu(x) \bar \varepsilon \gamma_\mu \lambda(x) \\ 
 &&  - a g \bar \varepsilon \gamma_\mu \lambda(x + a \hat\mu) U_\mu(x) \\
\delta \lambda(x) & = & -
\frac{i}{g} \sigma_{\rho\tau} {\cal G}_{\rho \tau}(x) \varepsilon 
\eneqa
where ${\cal G}_{\rho \tau}$ is the clover plaquette operator.

\subsection{Ward identities and operator mixing}

Compared to the numerical simulations \cite{federico}
a lattice perturbative calculation of the SUSY WIs introduces new aspects.
In order to do perturbation theory we have to fix the gauge, which
implies that new terms appear in the SUSY WIs:
the gauge fixing term (GF) and 
the Faddeev-Popov term (FP) while contact terms (CT) appear
off-shell \cite{dewit}.
Taking into account all contributions coming from the action, 
the bare SUSY WIs reads
$$
\left< {\cal O} \Delta_\mu S_\mu(x)-
2 m_0 {\cal O} \chi(x) + \left. \frac{\delta{\cal O}}
{\delta \bar \varepsilon(x)}\right|_{\varepsilon = 0} \right.
$$
\bma
\left. \left. - {\cal O} \frac{\delta S_{GF}}{\delta \bar \varepsilon(x)}
\right|_{\varepsilon = 0} - \left. {\cal O} \frac{\delta S_{FP}}
{\delta \bar \varepsilon(x)}\right|_{\varepsilon = 0} \right> =
\left< {\cal O} X_S(x) \right> \, .
\ema
$X_S$ is the symmetry breaking term, whose specific form depends on the 
choice of the lattice supercurrent. We define the lattice supercurrent to be
\bma
S_\mu(x) = - \frac{2 i}{g} \mbox{Tr} \left\{ {\cal G}_{\rho \tau}(x)
           \sigma_{\rho \tau} \gamma_\mu \lambda(x)  \right\} \, .
\ema
We choose a non-gauge invariant operator insertion  
${\cal O} := A_\alpha^a(y)\, \bar \lambda^b(z)$.
Gauge dependence implies that operator mixing with non-gauge 
invariant terms has to be taken into account for the 
operator renormalization.
$X_S$ mixes with operators of equal or lower dimension \cite{bochicchio}
\beeqa
X_S(x) &= & \bar{X}_S(x) - (Z_S - 1) \Delta_\mu S_\mu(x) - \\ 
&& 2 \tilde{m} \chi(x) - Z_T \Delta_\mu T_\mu(x) - \sum_i Z_{A_i} A_i \, .
\eneqa
The additional operators $A_i$ do not appear in the on-shell gauge invariant
numerical approach of \cite{federico}.
They are either BRS-exact, $A = \delta_{BRS} \tilde A$, or 
vanish using the equation of motion.
Moreover, because the $A_i$ do not appear in the SUSY WIs at tree level,
the $Z_{A_i}$ are $O(g^2)$.

We are forced to go off-shell, contrary to the numerical
simulations that are in the on-shell regime, otherwise it is
not possible to separate the contributions of $T_\mu$ and $S_\mu$.
Finally, infrared divergences are treated using the Kawai 
procedure \cite{kawai}, which gives
the general recipe to renormalize a Feynman diagram at one-loop order.

\subsection{Perturbative calculation}

We calculate the matrix elements of the terms in the SUSY WIs
for general external momenta $p$ (for the gauge field)
and $q$ (for the fermion field) and the projections over
$\gamma_\mu$ and $\gamma_\mu \gamma_5 $ matrices.

The lattice SUSY transformations of the gauge field $A_\mu$ 
are not identical to the continuum ones: the 
transformation of the gauge link $U_\mu$ determines the transformation 
properties of $A_\mu$. Up to $O(g^2)$,
\beeqa
&& \hspace{-0.6 cm} \delta A_\mu^b = i (\bar \varepsilon(x) \gamma_\mu \lambda^b(x) +
\bar \varepsilon(x + a \hat\mu) \gamma_\mu \lambda^b(x + a \hat\mu) ) 
\hspace{-0.1 cm}+ \\
&& \hspace{-0.7 cm} \frac{i}{2} a g f_{abc} 
(\bar \varepsilon(x) \gamma_\mu  \lambda^c(x) - \bar \varepsilon(x + a \hat\mu) 
\gamma_\mu \lambda^c(x + a \hat\mu)) A_\mu^a \\
&& - \frac{i}{24} a^2 g^2 (2 \delta_{ab} \delta_{cd} 
- \delta_{ac} \delta_{bd} - \delta_{ad} \delta_{bc} ) 
A_\mu^c A_\mu^d \, \bigg\{ \\
 && ( \bar \varepsilon(x) \gamma_\mu \lambda^a(x) + \bar \varepsilon(x + a \hat\mu) 
\gamma_\mu \lambda^a(x + a \hat\mu) ) \, \bigg\} \, ,
\eneqa
which reduces to the continuum SUSY transformation,
$ \delta A_\mu^a = 2 i \bar \varepsilon \gamma_\mu \lambda^a $,
in the limit $a \to 0$. 

At one-loop order, three propagator integrals on the lattice are tabulated 
in \cite{pana} in terms of lattice constants plus the continuum counterparts:
\bma
C_{0; \mu; \mu \nu; \mu \nu \rho}(p,q) =
\frac{1}{\pi^2}\!\int\!\!d^4k \frac{1; k_\mu; k_\mu k_\nu; k_\mu k_\nu k_\rho}
{k^2 (k + p)^2 (k + q)^2} \, .
\ema
With the help of \cite{velt,ball} one can express 
$C_0(p,q)$, which for arbitrary external momenta $p$ and $q$
is a complicated expression in terms of Spence functions, as 
\beeqa
C_0(p,q) & =& \frac{1}{\Delta} ( \mbox{Li}_2(\frac{p \cdot q - \Delta}{q^2} )
 - \mbox{Li}_2(\frac{p \cdot q + \Delta}{q^2} ) \\
&& + \frac{1}{2} \log(\frac{p \cdot q - \Delta}{p \cdot q + \Delta} ) 
\log( \frac{(q-p)^2}{q^2} )) \, , 
\eneqa
where $\Delta^2 = (p \cdot q)^2 - p^2 q^2$.
The $C_\mu(p,q)$, $C_{\mu \nu}(p,q)$, $C_{\mu \nu \rho}(p,q)$ can be written
recursively in terms of 
scalar functions $p^2, q^2, p \cdot q$ and $ C_0(p,q)$ multiplying 
Lorentz components of the external momenta $p$ and $q$ \cite{ball}.
The general results for arbitrary external momenta $p$ and $q$ are very long
(sometimes up to 1000 terms). Therefore a small momenta expansion 
is required.

\subsection{Renormalization Constants}

One can write the matrix elements in the form
\bma
\big< {\cal O} \Delta_\mu S_\mu \big> = S_F(q) \cdot \Lambda_{\Delta S} 
   \cdot D(p) \cdot \delta_{ab} \, ,
\ema
where $S_F(q)$ and $D(p)$ are the full gluino and gluon propagators,
$\delta_{ab}$ is the color factor and 
$\Lambda_{\Delta S}$ is the matrix element with amputated external propagators.
For small momenta, $\Lambda_{\Delta S}$ yields at one-loop order
\beeqa
\Lambda_{\Delta S} & = &  
   2 (p - q)_\mu p_\nu \sigma_{\nu \alpha} \gamma_\mu ( 1 + T^S_S ) \\
&& + i (p - q)_\mu (\not p \delta_{\mu \alpha} -
              p_\mu \gamma_\alpha) T^S_T + \cdots \, ,
\eneqa 
where $T^S_S$ is the coefficient of the one-loop contribution which is
proportional to the tree level expression of $\Lambda_{\Delta S}$,
$T^S_T$ is the coefficient of the contribution proportional to the tree
level $\Lambda_{\Delta T}$.
The tree level matrix element of 
$\big< {\cal O} \Delta_\mu T_\mu \big>$ reads
\bma
\Lambda_{\Delta T}  \, = \, 
      i (p - q)_\mu (\not p \delta_{\mu \alpha} - p_\mu \gamma_\alpha) \, .
\ema
Collecting the various contributions in the WIs and expanding them in terms
of a basis of Lorentz-Dirac structures, the renormalization and mixing
constants can be obtained from the coefficients as
\beeqa
-(Z_S - 1)&=&T^S_S + T^{CT}_S + T^{\chi}_S + T^{GF}_S + T^{FP}_S \\
- Z_T&=&T^S_T + T^{CT}_T + T^{\chi}_T + T^{GF}_T + T^{FP}_T \, .
\eneqa
We consider an appropriate choice of projections, which for small momenta
and on tree level read,
\beeqa
\mbox{Tr}(\gamma_c \Lambda_{\Delta T}) & = & 
4 i (p_c p_\alpha - p^2 \delta_{c \alpha} - p_c q_\alpha) \\ 
&& + p \cdot q \delta_{c \alpha} )  \\
\mbox{Tr}(\gamma_c \gamma_5 \Lambda_{\Delta T})& = & 0 \\
\mbox{Tr}(\gamma_c \Lambda_{\Delta S}) & = & 
8 i (p_c p_\alpha - p^2 \delta_{c \alpha} - p_c q_\alpha \\
&& + p \cdot q \delta_{c \alpha})  \\
\mbox{Tr}(\gamma_c \gamma_5 \Lambda_{\Delta S}) & = & 
8 i p_\mu q_\rho \varepsilon_{\mu \rho \alpha c} \, ,
\eneqa
in order to extract the coefficients.
Typically the $T_T^{(.)}, T_S^{(.)}$ 
are constants plus logarithms depending on the external momenta.
Our preliminary results for $Z_S $ and $Z_T$ 
show a good agreement with the numerical data of \cite{federico}.

\section{OUTLOOK}

It is possible to study the SUSY WIs by means of lattice 
perturbation theory and to determine the coefficients 
$Z_T$ and $Z_S$ in the off-shell regime.

The contributions for all diagrams have been written down explicitly
for small external momenta.
Preliminary results are in accordance with Monte Carlo data, but
still we have to increase the precision of the numerical integrations
and to perform several checks before presenting the final results in 
a forthcoming publication.


{\bf Acknowledgements:}
We thank G.~Rossi, M. Stingl and M.~Testa for stimulating discussions.



\end{document}